\documentclass[twocolumn,showpacs,preprintnumbers]{revtex4}
\usepackage{amssymb}
\usepackage{amsfonts}
\usepackage{amsmath}
\usepackage{graphicx}
\usepackage{dcolumn}
\usepackage{bm}

\input{tcilatex}

\begin{document}

\title{Three "quantum" models of competition and cooperation in interacting\
biological populations and social groups}
\author{E. D. Vol}
\email{vol@ilt.kharkov.ua}
\affiliation{B. Verkin Institute for Low Temperature Physics and Engineering of the
National Academy of Sciences of Ukraine 47, Lenin Ave., Kharkov 61103,
Ukraine.}
\date{\today }

\begin{abstract}
In present paper we propose the consistent statistical approach which
appropriate for a number of models describing both behavior of biological
populations and various social groups interacting with each other.The
approach proposed based on the ideas of quantum theory of open systems
(QTOS) and allows one to account explicitly both discreteness of a system
variables and their fluctuations near mean values.Therefore this approach
can be applied also for the description of small populations where standard
dynamical methods are failed. We study in detail three typical models of
interaction between populations and groups: 1) antagonistic struggle between
two populations 2) cooperation ( or, more precisely, obligatory mutualism)
between two species 3) the formation of coalition between two feeble groups
in their conflict with third one that is more powerful . The models
considered in a sense are mutually complementary and include the most types
of interaction between populations and groups. Besides this method can be
generalized on the case of more complex models in statistical physics and
also in ecology, sociology and other "soft' sciences.
\end{abstract}

\maketitle

\section{Introduction}

Among various classical open systems of interest in physics and also in
so-called "soft" sciences such as ecology,sociology,economics and so on, the
important role is played by the systems whose states in accordance with the
sense of the problem are specified by the set of integer variables $\left\{
n_{i}\right\} $ (where $n_{i}=0,1,2,...$, and $i=$1,2,....N , where N is a
number of degrees of freedom).For example in statistical physics $n_{i}$ may
represent occupation numbers of various cells in phase space of the nonideal
gas ,in ecology - numbers of individuals in distinct populations living in
the area and interacting with each other, in economics the number of
different companies operating on the market at the same time and so on.As
far as states of such systems change with time they are described as a rule
by a set of autonomous differential equations of the next form:%
\begin{equation}
\frac{dn_{i}}{dt}=F_{i}(\left\{ n_{\alpha }\right\} ),  \label{a1}
\end{equation}%
where $F_{i}(\left\{ n_{\alpha }\right\} )-$ some nonlinear state depending
functions determined by a concrete problem.

Obviously, that notation $\ $\ of \ Eq. (\ref{a1}) assumes that all
variables $\ n_{i\text{ }}$- are continuous quantities.In the case when all $%
n_{i}\gg 1$ such "smooth" approximation of discrete system is quite
reasonable.But for small occupation numbers when $n_{i}\gtrsim 1$, the
dynamical approach of Eq. (\ref{a1}) becomes inapplicable and it is
necessary to take into account both as discreteness of variables $n_{i}$ and
their possible fluctuations near mean values.The statistical method proposed
in the present paper has undoubted advantage compared with dynamic approach
Eq. (\ref{a1}) since it completely free from mentioned restrictions.

The method proposed is based on application of the quantum Lindblad master
equation (LME) for density matrix (DM) evolution of a open quantum Markov
system. The LME in general case has the next form( \cite{1s}, \cite{2s}):%
\begin{equation}
\frac{d\rho }{dt}=-\frac{i}{\hbar }\left[ H,\rho \right] +\sum\limits_{i}\ %
\left[ R_{i}\rho ,\text{ }R_{i}^{+}\right] +h.c,  \label{a2}
\end{equation}%
where H is some hermitian operator, describing the inner dynamics of open
system and $\left\{ R_{i}\right\} $ are a set of nonhermitian operators that
simulate different types of interaction of open system in question \ with
its environment).

Although Eq. (\ref{a2}) has obviously quantum origin, nevertheless, under
the certain conditions which we will discuss more detail later in this
paper, it may be applied also to the study of many classical open systems
,for example for the systems with discrete variables.The main advantage of
the LME compared with other similar master equations lies in the fact that
the number and form of operators $R_{i}$ entering in Eq. (\ref{a2}) for many
concrete problems can be defined from simple heuristic reasons. As soon as
such choice is made the consistent mathematical framework for the
description of the problem is at our disposal. Since the method proposed is
to some extent heuristic, we demonstrate its reasonableness and
effectiveness on the examples of different typical models which equally can
be applied to ecology for the description of behavior of interacting
populations and in sociology for studing dynamics of cooperating or
competing groups. Although in the present paper we specially restricted our
choice only the simplest examples, which admit the complete qualitative
analysis , it is clear that the method proposed can be generalized on more
complex cases without any difficulties of principle.The rest of the paper is
organized as follows.In Sect.2 we present the main features of approach
proposed , and discuss all conditions and restrictions for its application
.In Sect.3 we study the model of the struggle for existence between two
antagonistic biological populations in the case when total size of both
populations remains invariable.In Sect.4 we consider the model of
cooperation (or more precisely the model of obligatory mutualism) between
two distinct populations or social groups.In Sect.5 we study somewhat more
complicated model of three groups interaction in situation when two feeble
groups join together to confront successfully more powerful rival. In
conclusion we are summing up all results obtained in the paper.

Now let us turn to the concrete presentation of the paper.

\section{Description of the method}

As we already note the mathematical framework of the method proposed is the
Lindblad master equation (\ref{a2}) that is used in QTOS for the description
of open quantum Markov systems.Although this equation has undoubtedly
quantum origin under certain conditions it can be applied \ also for
statistical description of some classical open systems.

This significant point should be explained more detail. First of all note
that density matrix (DM) of quantum system apart from quantum correlations
containes also exhaustive information about its classical correlations. We
believe that information about classical correlations is recorded mainly in
diagonal elements of DM. Let us assume now that the open system of interest
is such that the relevant LME implies closed and complete set of equations
connecting only diagonal elements of its DM . Clearly, in this case the LME
could serve as appropriate framework for the statistical description of
classical analog of corresponding quantum system. Although similar situation
is realized for several classes of open classical systems, but in the
present paper we are interested in only special such class, namely classical
systems, whose states can be specified by integer variables $n_{i}$. Let us
describe this class more explicitly. Our first and main assumption is that
"hamiltonian term" in the r.h.s. of the LME (\ref{a2}) is missing.Besides we
demand that all operators $R_{i}$ in Eq. (\ref{a2}) have monomial form that
is: $R_{i}=C_{i}\prod\limits_{\alpha ,\beta }\left( a_{\alpha }^{+}\right)
^{k_{i,\alpha }}\left( a_{\beta }\right) ^{l_{i,\beta }}$ (in general case $%
C_{i\text{ }}$can be considered as functions of occupation numbers , but in
what follows they are assumed to be constants). In such situation one can
directly verify that all equations for diagonal elements of $\rho \left(
n_{1},...n_{N}\right) $, (where $n_{i}=a_{i}^{+}a_{i}$\ and $a_{i},a_{k}^{+}$
are usual Bose-operators, that is $\left[ a_{i},a_{k}^{+}\right] =\delta
_{ik},$) form complete set of differential-difference equations,which give
self-consistent description of the system.It should be note also that with
assumptions made the LME (\ref{a2}) actually reduced to the specific form of
Pauli master equation (PME) \cite{3s}, namely:%
\begin{equation}
\frac{d\rho _{n}}{dt}=\sum\limits_{n_{1}}(W_{nn_{1}}\rho
_{n_{1}}-W_{n_{1}n}\rho _{n}).  \label{a3}
\end{equation}

We could suggest also that for the" good enough" form of transition
probabilities $\ W_{nn_{1}}$ entering in the Eq. (\ref{a3}) one can find an
appropriate set of operators $R_{i}$ of monomial form such that the LME (\ref%
{a2})\ (without "hamiltonian term") and the PME (\ref{a3}) actually will
coincide. Now to complete the description of the method we need to specify
how one must choose a set of operators $\left\{ R_{i}\right\} $ for a
concrete problem of interest. The quality of the method proposed in our
opinion lies exactly in this point becouse this choice can be done according
to simple heuristic considerations.Namely, we believe the number and form of
a set $\left\{ R_{i}\right\} $ are entirely determined by the condition what
types of transitions $W_{nn_{1}}$ one want to take into account for the
concrete problem. The best way to illustrate all features of the method
proposed is to apply it to study various concrete models .Let us now turn to
this matter.

\section{Antagonistic struggle between two competing populations}

As the first example that illustrates all features of the method proposed we
consider the model of antagonistic opposition between two populations or
social groups struggling with each other for certain resourses or some other
preferences. In most naked form such struggle is realized when two tribes or
kins of cannibals living side by side in the area in a literal sense eat
each other ( but with unequal voracity). Assume for the simplicity that the
total number of individuals in both populations remains constant in this
struggle. As regards to sociology it could be for example the struggle
between two political\ parties competing at the parlament elections when the
total number of vacant seats is fixed. It is clear that in such situation
the benefit for one group necessarily means the failure for the other and
vice versa. Using the language of game theory, one can say that we consider
statistical model of two person game with zero sum (but from nonstandard
point of view). Let us turn now to the explicit mathematical formulation of
the model.We assume that our model of antagonistic struggle can be properly
described by the help of the LME with two operators: $R_{1}=\sqrt{\frac{a}{2}%
}a_{1}^{+}a_{2}$ and $R_{2}=\sqrt{\frac{b}{2}}a_{2}^{+}a_{1} $ (where
operator $R_{1\text{ }}$corresponds to events when first population or group
benefits and the second fails, and $R_{2}$ respectivly to the opposite one,
coefficients $a$ and $b$ reflect corresponding competitiveless of both
groups). After these principal assumptions remainder of our analysis is
completely rigorous and can be represented as the sequence of three
consecutive steps.

Step1. We write down the LME for the DM of the system , that characterize
total correlations between two groups of individuals with antagonistic
interaction. The equation has the next form:%
\begin{widetext}
\begin{equation}
\frac{d\rho }{dt}=a\left( a_{1}^{+}a_{2}\rho
a_{2}^{+}a_{1}-a_{2}^{+}a_{1}a_{1}^{+}a_{2}\rho \right) +b\left(
a_{2}^{+}a_{1}\rho a_{1}^{+}a_{2}-a_{1}^{+}a_{2}\rho a_{2}^{+}a_{1}\right) .
\label{a4}
\end{equation}
\end{widetext}One can easily see that Eq.(\ref{a4}) implies the closed set
of equations for diagonal elements of DM that describes only classical
correlations of the model in question, namely:%
\begin{widetext}
\begin{equation}
\frac{d\rho _{n_{1}n_{2}}}{dt}=a\left[ n_{1}\overline{n_{2}}\rho _{%
\underline{n_{1}}\overline{n_{2}}}-n_{2}\overline{n_{1}}\rho _{n_{1}n_{2}}%
\right] +b\left[ n_{2}\overline{n_{1}}\rho _{\overline{n_{1}}\underline{n_{2}%
}}-n_{1}\overline{n_{2}}\rho _{n_{1}n_{2}}\right] .  \label{a5}
\end{equation}
\end{widetext}(note that we use the notation $\overline{n}\equiv n+1,$ and $%
\underline{n}\equiv n-1$ to reduce the length of formula (\ref{a5})

Step2. It is convinient to represent the difference-differential system of
equations (\ref{a5}) in the form of equivalent differential equation. Such
representation can be obtained with the help of the generation function for
the distribution $\rho _{n_{1}n_{2}}$. According to definition, the
generating function (GF) \ $G\left( u,v,t\right) \equiv
\sum\limits_{n_{1}n_{2}}\rho _{n_{1}n_{2}}u^{n_{1}}v^{n_{2}}$. The
normalization condition $\sum\limits_{n_{1}n_{2}}\rho _{n_{1}n_{2}}=1$
implies that $G\left( 1,1,t\right) \equiv 1$. All moments of distribution $\
\rho _{n_{1}n_{2}}$ can be found by differentiating $G\left( u,v,t\right) $.
For example: $\overline{n_{1}}=\frac{\partial G}{\partial u}\mid _{u=v=1},%
\overline{n_{1\text{ }}(n_{1}-1)}=\frac{\partial ^{2}G}{\partial u\partial v}%
\mid _{u=v=1}$and so on.

One can verify directly that Eq. (\ref{a5}) implies the next equivalent
equation for GF $G\left( u,v,t\right) $:%
\begin{equation}
\frac{\partial G}{\partial t}=a\left( u-v\right) \frac{\partial ^{2}\left(
uG\right) }{\partial u\partial v}+b\left( v-u\right) \frac{\partial
^{2}\left( vG\right) }{\partial u\partial v}.  \label{a6}
\end{equation}%
Let us show that Eq. (\ref{a6}) admits an exhaustive qualitative analysis
because its general solution can be represented in the form of the
decomposition:%
\begin{equation}
G\left( u,v,t\right) =\sum\limits_{N}K_{N\text{ }}G_{N}\left( u,v,t\right) ,
\label{a7}
\end{equation}%
where $G_{N}\left( u,v,t\right) =A_{0}u^{N}+A_{1}u^{N-1}v+...A_{N}v^{N}$ is
normalized homogeneous polynomial of degree $N$ in variables $u$ and $v$;
and $\left\{ K_{N}\right\} $ is a set of constants satisfying to the
condition : $\sum\limits_{N}K_{N}=1.$The capability of decomposition Eq. (%
\ref{a7}) is a consequence of two reasons: 1) the linearity of Eq. (\ref{a6}%
) and 2) the existence of conservation law, namely, one can easily see that
total number of individuals in both populations $N=\overline{n_{1}}+%
\overline{n_{2}}.$ remains constant.

Thus we conclude that for the study of general dynamics of Eq. (\ref{a6})
enough to consider it in each of the subspaces with fixed $N$, \ where\ this
dynamics can be reduced to the simple system of linear equations for the
coefficients $A_{i}$ of homogeneous polynomial $G_{N}$:%
\begin{equation}
\frac{dA_{i}}{dt}=L_{ik}A_{k}.  \label{a8}
\end{equation}

In the Eq. (\ref{a8}) $L_{ik}$ is some nonhermitian $N+1\times N+1$ matrix
which can be easily calculated from Eq. (\ref{a6}) for any concrete $N$. For
example in the simplest case $N=2$ matrix $L_{ik}$ has the form: $L_{2}=%
\begin{pmatrix}
-2b & 2a & 0 \\ 
2b & -\left( 2a+2b\right) & 2a \\ 
0 & 2b & -2a%
\end{pmatrix}%
$. Having in hands decomposition (\ref{a7}) we are able to answer on the
main question: how the time evolution of arbitrary initial distribution $%
G_{0}\left( u,v\right) =G_{N}\left( u,v,t=0\right) $ is happened? To answer
this question let us begin with the end and find the stationary solutions of
Eq. (\ref{a6}). Equating \ the r.h.s. of (\ref{a6}) to zero we obtain that:%
\begin{equation}
G^{st}\left( u,v\right) =\frac{A\left( u\right) +B\left( v\right) }{au-bv}.
\label{a9}
\end{equation}%
To obtain required polynomial form of $G\left( u,v,t\right) $ we must choose 
$A_{N}\left( u\right) $ as $C_{N}\left( au\right) ^{N+1}$and $B\left(
v\right) $ as $-C_{N}\left( bv\right) ^{N+1}$ after that Eq. (\ref{a9})
leads to the next result:%
\begin{equation}
G_{N}^{st}\left( u,v\right) =C_{N}\left[ \left( au\right) ^{N}+\left(
au\right) ^{N-1}\left( bv\right) +...\left( bv\right) ^{N}\right] ,
\label{a10}
\end{equation}%
( where $C_{N}=a^{N}+a^{N-1}b+...b^{N}$ is normalization constant). With the
help of Eq. (\ref{a10}) for the GF one can find all moments that is all
statistical characteristics of the system of interest in its stationary
states. In this connection we would like to mention one curious result.Let N
is total ( and conserved) number of individual in both populations and
assumed that $a\geq b$.Then with the help of (\ref{a10}) one can show that
when $t$ tends to infinity, the fraction $\frac{\overline{n_{2}}}{\overline{%
n_{1}}}$ tends to zero, that is less voracious population gradually
disappears.

This conclusion is completely coincides with result obtained from dynamical
approach of Eq. (\ref{a1}), since for $N\gg 1,$ the role of discreteness of
variables $n_{i}$ and their fluctuations becames negligible. On the other
hand the expression (\ref{a10}) is valid also in the case of small
populations where dynamical approach is not applicable. For example when $%
N=2 $ \ GF is equal to: $G_{2}=\frac{a^{2}u^{2}\text{ }+\text{ }abuv\text{ }%
+b^{2}v^{2}}{a^{2}+ab+b^{2}}$ and we can find that $\overline{n_{1}}=\frac{%
2a^{2}+ab}{a^{2}+ab+b^{2}};$ $\ \overline{n_{2}}=\frac{2b^{2}+ab}{%
a^{2}+ab+b^{2}}$, $\overline{n_{1}n_{2}}=\frac{ab}{a^{2}+ab+b^{2}}$. Going
over to the study of time evolution of arbitrary initial distribution $%
G_{0}\left( u,v\right) $ we propose the following precription: using Eq. (%
\ref{a7}) one must expand $G_{0}$ in homogeneous polynomials $G_{N}$, thus
determining a set of $K_{N}$ , and after that we can immediately write down
the definitive result, namely $G\left( u,v,\infty \right)
=\sum\limits_{N}K_{N}G_{N}^{st}\left( u,v\right) .$In what follows we will
call states with GF$\ G_{N}$ \ the pure states.

Note also that using the Euler theorem about homogeneous functions: $u\frac{%
\partial G_{N}}{\partial u}+v\frac{\partial G_{N}}{\partial v}=NG_{N}$ , one
can easily obtain for the pure states two simple relations:%
\begin{equation}
\overline{n_{1}^{2}}+\overline{n_{1}n_{2}}=N\overline{n_{1\text{ }}}\text{
and }\overline{n_{2}^{2}}+\overline{n_{1}n_{2}}=N\overline{n_{2}}.
\label{a11}
\end{equation}%
If we define by standard way the variance of the number of individuals in
every population as $\sigma _{i}=\overline{n_{i}^{2}}-\left( \overline{n_{i}}%
\right) ^{2}$ (where i=1,2) and correlation coefficient $k=\frac{\overline{%
n_{1}n_{2}}-\overline{n_{1}}\text{ }\cdot \text{ }\overline{n_{2}}}{\sqrt{%
\sigma _{1}\sigma _{2}}}$then the Eq. (\ref{a11}) implies that $\sigma
_{1}=\sigma _{2}\equiv \sigma =\overline{n_{1}}$ $\overline{n_{2}}-\overline{%
n_{1}n_{2}}$ \ and hence for any values $a$ and $b$ correlation coefficient $%
k$ in any \ pure state is equal to $-1.$This result exactly justifies the
name for considered model as antagonistic model of interaction between two
populations or groups.

\section{Cooperation between two populations or social groups}

In this part we consider the model of interaction between two groups of
individuals in which both populations either together get the benefit in the
struggle for existence or under adverse circumstances are failing
together.In biology such kind of interaction is known as symbiosis( or more
precisely as obligatory mutualism).As the simplest example of such
interaction we can specify the alliance between hermit- crab and actinium,
that is sea coral.$\left[ 4\right] $Actinium attached to the shell in which
crab lives , and then moves with it and eats the remaines of its food. On
the other hand actinium protects the crab from its enemies with special
cells located in its tentacles.Thus neither actinium neither hermit -crab
cannot successfully exist without each other.As regards to sociology the
number of distinct groups that share common goals and (or) values and
cooperates with each other to achieve them are truly unlimited.To formulate
the mathematical model describing the case of two groups obligatory
cooperation we start again \ from Lindblad equation (\ref{a2}) , but now to
accordance with the sence of the problem as a collection of operators $R_{i}$
we choose two operators:$R_{1}=\sqrt{\frac{a}{2}}a_{1}^{+}a_{2}^{+}$ and $%
R_{2}=\sqrt{\frac{b}{2}}a_{1}a_{2}$ .The first operator describes the events
in which both groups get the benefit while the second one is responsible for
the common failure.The required Lindblad equation for diagonal elements of
density matrix of the system $\rho _{n_{1}n_{2}}$ in this case takes the
form:%
\begin{widetext}
\begin{equation}
\frac{\partial \rho _{n_{1}\text{ }n_{2}}}{\partial t}=a\left[
n_{1}n_{2}\rho _{\underline{n_{1}}\text{ }\underline{n_{2}}}-\overline{n_{1}}%
\cdot \overline{n_{2}}\cdot \rho _{n_{1}n_{2}}\right] +b\left[ \overline{%
n_{1}}\cdot \overline{n_{2}}\cdot \rho _{\overline{n_{1}}\text{ }\overline{%
n_{2}}}-n_{1}n_{2}\rho _{n_{1}\text{ }n_{2}}\right] .  \label{a12}
\end{equation}
\end{widetext}(In the Eq. (\ref{a12}) as before we used the notation:$%
\overline{n_{i}}=n_{i}+1$ and $\underline{n_{i}}=n_{i}-1)$. It is easy to
verify that difference-differential equation (\ref{a12}) by introducing the
generating function $G\left( u,v,t\right) =\sum\limits_{n_{1}n_{2}}\rho
_{n_{1}n_{2}}u^{n_{1}}v^{n_{2}}$ can be transformed to the equivalent form
of single \ \ differential \ equation , namely:%
\begin{equation}
\frac{\partial G}{\partial t}=\left( 1-uv\right) \frac{\partial ^{2}}{%
\partial u\partial v}\left[ \left( b-auv\right) G\right] .  \label{a13}
\end{equation}%
Note that Eq. (\ref{a13}) as well as Eq. (\ref{a16}) admits the exhaustive
qualitative analysis, due to linearity and the presence of conservation law.
We begin our analysis of Eq. (\ref{a13}) with finding its stationary
solutions. Equating the r.h.s. of Eq. (\ref{a13}) to zero, we obtain that a
set of stationary solutions of Eq. (\ref{a13}) can be represented in the
form:%
\begin{equation}
G_{st}\left( u,v\right) =\frac{A\left( u\right) +B\left( v\right) }{b-auv},
\label{a14}
\end{equation}%
where functions $A\left( u\right) $ and $B\left( v\right) $ yet to be
determined. As the second step let us write down equations of motion for the
occupation numbers mean values of both populations.By differentiaiting of
Eq. (\ref{a13}) with respect to $u$ and $v$ respectively and then putting $%
u=v=1$ we obtain:%
\begin{eqnarray}
\frac{\partial \overline{n_{1}}}{\partial t} &=&\frac{\partial \overline{%
n_{2}}}{\partial t}  \label{a15} \\
\frac{\partial \overline{n_{2}}}{\partial t} &=&a\left( \overline{n_{1}}+%
\overline{n_{2}}+1\right) +\left( a-b\right) \overline{n_{1}n_{2}}.  \notag
\end{eqnarray}%
In what follows we consider the case $b\geq a.$ The Eq. (\ref{a15}) implies
that $\overline{n_{1}}-\overline{n_{2}}=N$ is the integral of motion for Eq.
(\ref{a13})$.$ Further we will consider $N$ as a fixed integer(positive,zero
or negative).Let us prove now that the dynamics of Eq. (\ref{a13}) can be
considered as superposition of independent dynamics for each $N$ \
respectively .\ Indeed, let $N$ is some positive integer and take as
generating function the normalized function of the form: $G_{N}\left(
u,v,t\right) =u^{N}f\left( uv,t\right) $ (where $f\left( 1,t\right) =1$). It
is easy to see that for such \ GF \ $\ \overline{n_{1}}-\overline{n_{2}}=N$
\ for arbitrary $\ f\left( x,t\right) \equiv f\left( uv,t\right) $.
Substituting this expression for GF in Eq. (\ref{a13}) after simple algebra
we obtain the closed equation of evolution for the function $f\left(
x,t\right) $:%
\begin{widetext}
\begin{equation}
\frac{\partial f\left( x,t\right) }{\partial t}=\left( 1-x\right) \left[
\left( N+1\right) \frac{\partial }{\partial x}\left( b-ax\right) f+x\frac{%
\partial ^{2}}{\partial x^{2}}\left( b-ax\right) f\right] .  \label{a16}
\end{equation}
\end{widetext}Let us prove now that $f\left( x,t\right) $ tends to $\frac{C}{%
b-ax}$ when t tends to infinity.For this purpose we introduce auxillary
function $h\left( x,t\right) =\left( b-ax\right) f\left( x,t\right) $. Then
Eq. (\ref{a13}) implies the next equation for function $h\left( x,t\right) $:%
\begin{eqnarray}
\frac{\partial h}{\partial t} &=&\left( 1-x\right) \left( b-ax\right) \left[
\left( N+1\right) \frac{\partial h}{\partial x}+x\frac{\partial ^{2}h}{%
\partial x^{2}}\right] \equiv  \notag \\
&\equiv &-\frac{\left( 1-x\right) \left( b-ax\right) }{x^{N}}\frac{\delta F%
\left[ h\left( x\right) \right] }{\delta h},  \label{a17}
\end{eqnarray}%
where $F\left[ h\left( x\right) \right] $ is the functional of $h\left(
x\right) $ which has the form: $F\left[ h\left( x\right) \right] =\frac{1}{2}%
\int\limits_{0}^{1}x^{N+1}\left( \frac{\partial h}{\partial x}\right) ^{2}dx$%
; The Eq. (\ref{a17}) implies that $\frac{dF}{dt}=-\int\limits_{0}^{1}\left(
1-x\right) \left( b-ax\right) \left( \frac{\delta F}{\delta h}\right)
^{2}dx\leq 0$ and hence when $t$ tends to infinity functional $F\left[
h\left( x,t\right) \right] $ tends to zero and therefore $h\left( x,t\right) 
$ tends to certain constant value (remind that we assume $b\geq a$).QED.It
is clear that all above arguments remain valid also in \ \ the case when
integer $N\leq 0,$ if one take as generating function $G_{N}=v^{\mid N\mid
}g\left( uv,t\right) $ .Thus we reach the required conclusion:the general
solution of Eq. (\ref{a13}) admits the decomposition: $G\left( u,v,t\right)
=\sum\limits_{N=-\infty }^{N=\infty }K_{N}$ $G_{N}\left( u,v,t\right) $ . It
means that for the study of evolution of arbitrary initial GF $G_{0}=G\left(
u,v,t=0\right) $ enough to consider separately dynamics of each it
component. Also we have proved that any component of initial GF of the form $%
u^{N}f_{0}\left( uv\right) $ (such GF we again connect with certain pure
state) transforms to the distribution $\frac{u^{N}\left( b-a\right)
f_{0}\left( 1\right) }{\left( b-auv\right) }.$Similar result is valid for
the initial GF component $\ $of the form $v^{M}g_{0}\left( uv\right) .$Thus
qualitative description of evolution for arbitrary initial GF $G_{0}$ is
completed. In conclusion of this part we present two notable results
relating to evolution of pure states in this model.Result1.Let $%
G_{0}=u^{N}f\left( uv\right) $ .We have just proved that at any time $%
G\left( u,v,t\right) $ has the same form and hence satisfies to the equation:%
\begin{equation}
u\frac{\partial G}{dt}-v\frac{\partial G}{dt}=NG.  \label{a18}
\end{equation}%
The Eq. (\ref{a18}) implies relations which are similar to the relations (%
\ref{a11}) for foregoing model,namely:

\begin{eqnarray}
\overline{n_{1}}-\overline{n_{2}} &=&N,\   \notag \\
\overline{n_{1}^{2}}-\overline{n_{1}n_{2}} &=&N\overline{n_{1}},  \label{a19}
\\
\overline{n_{2}}-\overline{n_{1}n_{2}} &=&N\overline{n_{2}},  \notag
\end{eqnarray}%
from which we are able to obtain expressions for variations of statistical \
distributions of $n_{1}$ and $n_{2},$ namely: $\sigma _{1}=\overline{%
n_{1}^{2}}-\left( \overline{n_{1}}\right) ^{2}$ $=\sigma _{2}$ $=\overline{%
n_{2}^{2}}-\left( \overline{n_{2}}\right) ^{2}$ $=$ $\overline{n_{1}n_{2}}-%
\overline{n_{1}}\overline{n_{2}}$ and hence correlation coefficient $k$
between two populations 1 and 2 is : $k_{12}=\frac{\overline{n_{1}n_{2\text{ 
}}}-\text{ \ }\overline{n_{1}}\text{ }\overline{n_{2}}}{\sqrt{\sigma
_{1}\sigma _{2}}}=1.$Thus we come to the conclusion that all pure states in
this model are maximally correlated.

Result 2. Let us again assume that initial pure state has GF of the form: $%
G_{0}=u^{N}f_{0}\left( uv\right) $. Then mean occupation numbers of two
populations are: $\overline{n_{1}}=N+\frac{\partial f_{0}}{\partial x}\mid
_{x=1}$and $\overline{n_{2}}=\frac{\partial f_{0}}{\partial x}\mid _{x=1}.$

When $t$ tends to infinity then $G$ tends to $G_{{}}^{st}=\frac{\left(
b-a\right) u^{N}}{b-auv}$ and occupation numbers in stationary state are: $%
\overline{n_{1}}=N+\frac{a}{b-a}$ and $\overline{n_{2}}=\frac{a}{b-a}$. Let $%
\frac{\partial f_{0}}{\partial x}\ll 1$ then initial number $\overline{n_{2}}
$ is very small,but in situation when ratio $\varkappa =\frac{a}{b}$ tends
to one, the number of individuals in both populations increases with time
without any bound. However, the difference between occupation numbers\ $%
\overline{n_{1}}$ and $\overline{n_{2}}$ \ remains constant as before.

\section{The formation of coalition between two populations or social groups
in the struggle against common rival}

In this part we consider more complex model of interaction between three
distinct populations which is, in a sence, the superposition of two
foregoing models.

We will focus on the situation when two feeble groups of individuals enter
into an alliance to resist together the common rival.Thus in this model on
the one hand there is a cooperation between two feeble groups but on the
other hand there is mutual antagonism in relation to common
enemy.Unfortunately, I find it difficult to bring vivid examples of such
coalition in biology (although they are apparently exist).On the other hand
it is clearly that in social, political and military conflicts there are the
great number of forming such coalitions. However, I include this example in
the present paper mainly to demonstrate how one can construct more and more
complex models based on simple ones.

Let us now turn to the explicit mathematical formulation of this model .As
before we are starting from the Lindblad equation (\ref{a2}) but this time
we choose as the set of $R_{i}$ the next two operators: $R_{1}=\sqrt{\frac{a%
}{2}}a_{1}^{+}a_{2}a_{3}$ and $R_{2}=\sqrt{\frac{b}{2}}%
a_{1}a_{2}^{+}a_{3}^{+}$. We believe that operator $R_{1}$ describes events
when more stronger population with index 1 wins in the conflict and
respectively the coalition formed from two populations with indices 2 and 3
fail, while the operator $R_{2}$ takes into account the contrary events.The
Lindblad equation for the diagonal elements of density matrix $\rho
_{n_{1}n_{2}n_{3\text{ }}}$in this case takes the form: 
\begin{widetext}
\begin{equation}
\frac{\partial \rho _{n_{1\text{ }}n_{2}\text{ }n_{3}}}{\partial t}=a\left[
n_{1}\cdot \overline{n_{2}}\cdot \overline{n_{3}}\rho _{\underline{n_{1}}%
\text{ }\overline{n_{2}}\text{ }\overline{n_{3}}}-\overline{n_{1}}\cdot
n_{2}\cdot n_{3}\rho _{n_{1}\text{ }n_{2}\text{ }n_{3}}\right] +b\left[ 
\overline{n_{1}}\cdot n_{2}\cdot n_{3}\rho _{\overline{n_{1}}\text{ }%
\underline{n_{2}}\text{ }\underline{n_{3}}}-n_{1}\cdot \overline{n_{2}}\cdot 
\overline{n_{3}}\rho _{n_{1}\text{ }n_{2}\text{ }n_{3}}\right] .  \label{a20}
\end{equation}
\end{widetext}( we again use in Eq. (\ref{a20}) the notation: $\overline{%
n_{i}}=n_{i}+1$ and $\underline{n_{i}}=n_{i}-1$).

As before we convert the system of difference-differential equations Eq. (%
\ref{a20}) to equivalent differential equation for generation function : $%
G\left( u,v,w,t\right) =\sum\limits_{n_{1},n_{2},n}\rho
_{n_{1}n_{2}n_{3}}u^{n_{1}}v^{n_{2}}w^{n_{3}}$. Required equation for GF in
this model reads as:%
\begin{equation}
\frac{\partial G}{\partial t}=\left( u-vw\right) \frac{\partial ^{3}}{%
\partial u\partial v\partial w}\left[ \left( au-bvw\right) G\right] .
\label{a21}
\end{equation}%
The Eq. (\ref{a21}) implies the next equations of motion for mean occupation
numbers $\overline{n_{1}}$ ,$\overline{n_{2}}$ , $\overline{n_{3}}$:%
\begin{widetext}
\begin{eqnarray}
\frac{\overline{\partial n_{1}}}{\partial t} &=&a\overline{n_{2}n_{3}\left(
n_{1}+1\right) }-b\overline{n_{1}\left( n_{2}+1\right) \left( n_{3}+1\right) 
}  \notag \\
\frac{\overline{\partial n_{2}}}{\partial t} &=&\frac{\overline{\partial
n_{3}}}{\partial t}=-a\overline{n_{2}n_{3}\left( n_{1}+1\right) }+b\overline{%
n_{1}\left( n_{2}+1\right) \left( n_{3}+1\right) }  \label{a22}
\end{eqnarray}
\end{widetext}One can see that Eq. (\ref{a22}) imply the existence of two
integrals of motion, that can be written as:1) $\overline{n_{1}}+\overline{%
n_{2}}=N+p$ and 2) $\overline{n_{1}}+\overline{n_{3}}=N.$(we will consider
further only the case when $N$ and $p$ are positive integers). The presence
of these integrals allows one to give complete qualitative description of
evolution (\ref{a21}) for any initial generating function $G_{0}\left(
u,v,w,t=0\right) .$To realize this intention we firstly find the stationary
solutions of Eq. (\ref{a21}). Equating its r.h.s to zero we obtain required
result:%
\begin{equation}
G^{st}\left( u,v,w,\right) =\frac{A\left( u,v\right) +B\left( u,w\right)
+C\left( v,w\right) }{au-bvw},  \label{a23}
\end{equation}%
(where functions $A\left( u,v\right) ,B\left( u,w\right) ,C\left( v,w\right) 
$ must be determined in such a way to obtain for $G^{st}$ some polynomial
expansion.It is easy to see that after relevant choice we can rewrite Eq. (%
\ref{a23}) in the required polynomial form:%
\begin{widetext}
\begin{equation}
G_{{}}^{st}\left( u,v,w\right) =\left[ K\left( v\right) +L\left( w\right) %
\right] \left[ \left( au\right) ^{N}+\left( au\right) ^{N-1}\left(
bvw\right) +...\left( bvw\right) ^{N}\right] ,  \label{a24}
\end{equation}
\end{widetext}(where polynomials $K\left( v\right) ,L\left( w\right) $ are
determined by initial GF \ ). Now \ we will seek a complete set of pure
states with self-closed dynamics, on which arbitrary generating function can
be expanded.To this end let us consider the generating function in the form: 
$G_{p}=v^{p}f\left( u,vw,t\right) .$ Substituting this expression in Eq. (%
\ref{a21}) after simple algebra we obtain the closed evolution equation for
the function $f\left( u,x,t\right) \equiv f\left( u,vw,t\right) $, namely:%
\begin{widetext}
\begin{equation}
\frac{\partial f}{\partial t}=\left( u-x\right) \frac{\partial }{du}\left\{ x%
\frac{d^{2}}{dx^{2}}\left[ \left( au-bx\right) f\right] +\left( p+1\right) 
\frac{d}{dx}\left[ \left( au-bx\right) f\right] \right\} .  \label{a25}
\end{equation}
\end{widetext}It is easy to see that the general solution of Eq. (\ref{a25})
can be represented as some linear superposition of homogeneous polynomials
of $N$ degree in variables $x$ and $u$, that is $f\left( u,x,t\right)
=\sum\limits_{N}P_{N}\left( u,x,t\right) $

Besides it is clear, if we take the generation function in the form $%
G_{p}=w^{p}g\left( u,vw\right) $ we come to the similar result.Now we are
able to claim that the general solution of Eq. (\ref{a21}) can be expended
on these two classes of pure states.We can write this decisive for the
further study of this model result as:%
\begin{equation}
G\left( u,v,w,t\right) =\sum\limits_{N,p}\left[ v^{p}P_{N}\text{ }\left(
u,vw\right) +w^{p}R_{N}\text{ }\left( u,vw\right) \right] .  \label{a26}
\end{equation}%
(in Eq. (\ref{a26}) summation is over all integers $N$ and $p$, and $%
P_{N}\left( u,x\right) $ , $R_{N}\left( u,x\right) $ are homogeneous
polynomials of $N$ degree in $u$ and $x\equiv vw$ depending on $p$ as well).

The result obtained allows one to restrict the study of the Eq. (\ref{a21})
to the case of pure states evolution only. So, let us consider the pure
state which continually has the form: $\ \ \ \ \ \ G_{pN}\left(
u,v,w,t\right) =v^{p}R_{N}\left( u,vw,t\right) $ (where $R_{N\text{ \ }}$is
homogeneous polynomial of $N$ degree ).We specify here only two notable
properties of such states in this model.

1) Cooperating groups being in pure states are fully correlated,while
conflicting groups are fully anticorrelated.To prove this result note that
pure states in question satisfy to the equations:%
\begin{equation}
u\frac{\partial G_{pN}}{\partial u}+w\frac{\partial G_{pN}}{\partial w}%
=NG_{pN},  \label{a27}
\end{equation}%
and%
\begin{equation}
v\frac{\partial G_{pN}}{\partial v}-w\frac{\partial G_{pN}}{\partial w}%
=pG_{pN}.  \label{a28}
\end{equation}%
The Eq. (\ref{a27}) implies following relations:a) $\overline{n_{1}}+%
\overline{n_{3}}=N,$and b) $\overline{n_{1}}+\overline{n_{1}\left(
n_{1}-1\right) }+\overline{n_{1}n_{3}}=N\overline{n_{1}}$ from which one can
find the variance of first group population,namely: $\ \ \ \ \ \ \ \sigma
_{1}=\overline{n_{1}^{2}}-(\overline{n_{1}})^{2}=N\overline{n_{1}}-\left( 
\overline{n_{1}}\right) ^{2}-\overline{n_{1}n_{3}}=$ $\ \overline{n_{1}}\ 
\overline{n_{3}}-\overline{n_{1}n_{3}}$ .Selfsame result one can obtain for
the variance $\sigma _{3}$ and hence correlation coefficient $\ k_{13}=\frac{%
\overline{n_{1}n_{3}}-\overline{n_{1}}\text{ \ }\overline{n_{3}}}{\sqrt{%
\sigma _{1}\sigma _{3}}}$ between groups 1 and 3 is equal to $-1$.Similar
algebra results in that $k_{12}=-1$, and $k_{23}=1.$QED.Let us consider now
another important question: how the properties of stationary pure state $%
G_{pN}^{st}$ \ depend on coefficients $a$ and $b$ and also on values of
integers $N$ and $p$.We intend to analyze this problem in all details
elsewhere and in this paper, as an illustration, consider only single
special case when integer $N$ tends to infinity while integer $p$ remains
finite.So, let stationary GF is:%
\begin{equation}
G_{pN}^{st}=v^{p}\left[ \left( au\right) ^{N}+\left( au\right) ^{N-1}\left(
bvw\right) +....\left( bvw\right) ^{N}\right] .  \label{a29}
\end{equation}%
We want to understand how in this stationary state ratio of mean occupation
numbers in populations 1 and 3 depend on the coefficients $a$ and $b.$With
the help of GF from Eq. (\ref{a29}) the required ratio $\frac{\overline{n_{1}%
}}{\overline{n_{3}}}$ can be represented in the form:%
\begin{widetext}
\begin{equation}
\frac{\overline{n_{1}}}{\overline{n_{3}}}=\frac{Na^{N}+\left( N-1\right)
a^{N-1}b+.....ab^{N-1}}{Nb^{N}+\left( N-1\right) b^{N-1}a+.....a^{N-1}b}=%
\frac{N\varkappa ^{N}+\left( N-1\right) \varkappa ^{N-1}+...\varkappa }{%
N+\left( N-1\right) \varkappa +....\varkappa ^{N-1}}=\frac{\varkappa \frac{d%
}{d\varkappa }\left[ Lnf_{N}\left( \varkappa \right) \right] }{N-\varkappa 
\frac{d}{d\varkappa }\left[ Lnf\left( \varkappa \right) \right] }.
\label{a30}
\end{equation}
\end{widetext}(where we use the notation: $\varkappa =\frac{a}{b}$ and $%
f_{N}\left( \varkappa \right) =1+\varkappa +....+\varkappa ^{N}=\frac{%
\varkappa ^{N+1}-1}{\varkappa -1}$). Let $N$ tends to infinity. Using the
Eq. (\ref{a30}) it is easy to see that $\frac{\overline{n_{1}}}{\overline{%
n_{3}}}$tends to $\left( \frac{\varkappa -1}{\varkappa }\right) N$ \ when $%
\varkappa >1$, and $\frac{\overline{n_{1}}}{\overline{n_{3}}}$ tends to $%
\frac{\varkappa }{\left( 1-\varkappa \right) N}$ when $\varkappa <1$. Thus
in the case when $\varkappa >1$ we obtain: $\overline{n_{1}}\rightarrow N,%
\overline{n_{2}}\rightarrow p$ and $\overline{n_{3}}\rightarrow \frac{%
\varkappa }{\varkappa -1}$ while in the case $\varkappa <1$ we get :$%
\overline{n_{1}}\rightarrow \frac{\varkappa }{1-\varkappa },\overline{n_{2}}%
\rightarrow N+\left( p-\frac{\varkappa }{1-\varkappa }\right) $ and $%
\overline{n_{3}}\rightarrow N.$In other words if $N\gg 1$ and $\varkappa >1$
the first population becames dominating, contrary in the case $\varkappa <1$
coalition dominates.On the other hand when $N\simeq p$ the behavior of the
model becames considerably more complex and requires a separate study.

Let us briefly summing up the main results of present paper. Based on ideas
of QTOS we propose the consistent approach for statistical description of
open classical systems with integer variables. We proved that for broad
class of open systems possesing specific restrictions on the form of their
interaction with environment the LME actually reduced to the Pauli master
equation for diagonal elements of density matrix.This fact gives one the
reason to use the LME for quantitative study of different problems relating
both to statistical physics and to various "soft' sciences such as ecology,
sociology, economics and so on.

I would like to acknowledge L.A.\ Pastur for useful discussions of the
results of this paper.

\end{document}